\documentclass[twocolumn,tighten]{aastex63}
\submitjournal{ApJL}

\hypersetup{colorlinks,%
 citecolor=blue,%
 linkcolor=blue}
\newcommand{\kms}{km\,s$^{-1}${}}
\newcommand{\msun}{M$_{\sun}${}}
\shorttitle{A dwarf galaxy group in the local universe}
\shortauthors{Paudel et al.}

\newcommand{\y}[1]{\textcolor{red}{#1}}

\begin{document}

\title{Discovery of a Rare Group of Dwarf Galaxies in the Local Universe }

\author[0000-0003-2922-6866]{Sanjaya Paudel}
\affil{Department of Astronomy \& Center for Galaxy Evolution Research, Yonsei University, Seoul 03722, Republic Of Korea}

\author[0000-0002-5513-5303]{Cristiano G. Sabiu}
\affil{Natural Science Research Institute (NSRI), University of Seoul, Seoul 02504, Republic of Korea}

\author[0000-0002-1842-4325]{Suk-Jin Yoon}
\affil{ Department of Astronomy \& Center for Galaxy Evolution Research, Yonsei University, Seoul 03722, Republic Of Korea}
\email{sjyoon0691@yonsei.ac.kr}

\author[0000-0003-3343-6284]{Pierre-Alain Duc}
\affil{Universit\'e de Strasbourg, CNRS, Observatoire astronomique de Strasbourg, UMR 7550, F-67000 Strasbourg, France}

\author[0000-0002-6841-8329]{Jaewon Yoo}
\affil{Quantum Universe Center, Korea Institute for Advanced Study (KIAS), 85 Hoegiro, Dongdaemun-gu, Seoul 02455, Republic of Korea}

\author[0000-0003-4552-9808]{Oliver M{\"u}ller}
\affil{Laboratoire d’astrophysique, École Polytechnique Fédérale de Lausanne (EPFL), Observatoire, 1290 Versoix, Switzerland}


\begin{abstract}
We report the discovery of a rare isolated group of five dwarf galaxies located at z = 0.0086 ($D$ = 36 Mpc). 
All member galaxies are star-forming, blue, and gas-rich with $g-r$ indices ranging from 0.2 to 0.6 mag, and two of them show signs of ongoing mutual interaction. 
The most massive member of the group has a stellar mass that is half of the Small Magellanic Cloud stellar mass, and the median stellar mass of the group members is 7.87 $\times$ 10$^{7}$ \msun. 
The derived total dynamical mass of the group is $M_{\rm dyn}$ = 6.02$\times$10$^{10}$ \msun, whereas its total baryonic mass (stellar + HI) is 2.6$\times$10$^{9}$ M$_{\sun}$, which gives us the dynamical to baryonic mass ratio of 23.  
Interestingly, all galaxies found in the group are aligned along a straight line in the plane of the sky. 
The observed spatial extent of the member galaxies is 154 kpc, and their relative line-of-sight velocity span is within 75 \kms. 
Using the spatially resolved optical spectra provided by DESI EDR, we find that three group members share a common rotational direction. 
With these unique properties of the group and its member galaxies, we discuss the possible importance of such a system in the formation and evolution of dwarf galaxy groups and in testing the theory of large-scale structure formation.
\end{abstract}

\keywords{Unified Astronomy Thesaurus concepts: Interacting galaxies(802), Dwarf galaxies(416), Galaxy groups(597)}


\section{Introduction}

 The Lambda Cold Dark Matter ($\Lambda$CDM) model is successful in explaining the large-scale structure of the Universe, but it encounters difficulties with explaining many aspects of 
 dwarf galaxies \citep{White78,Frenk12,Bullock17}. According to the $\Lambda$CDM theory, even low-mass halos possess detectable substructures, which implies that dwarf galaxies ought to have sub-halos and form individual groups \citep{Diemand08,Wheeler15}. Consequently, simulations predict mergers involving only dwarf galaxies \citep{Deason14}.

However, statistical analysis of both observational and simulation data estimated that fewer than 5\% of dwarf galaxies have close companions, and there is less than a 0.004\% chance that a dwarf galaxy would be part of a quad group (with the number of members equal to 4) of dwarf galaxies \citep{Stierwalt17,Besla18}. Moreover, very few of these close companion dwarf galaxies could undergo a merger.

The interaction between the Large Magellanic Cloud (LMC) and the Small Magellanic Cloud (SMC) is a unique example that has been extensively studied in both observations and theoretical models \citep{Putman98,Besla10,Glatt10,Besla12,Kallivayalil13,Onghia16,Kallivayalil18}. In addition, recent deep observation has revealed several new satellites around the LMC-SMC system \citep{Koposov15,Wagner15,Koposov18}. The discovery of the LMC group has revitalized the importance of LMC-SMC on the evolution of the Milky-Way (MW) satellite system and the MW itself; particularly, it is hypothesized that a fraction of the MW satellites is accreted as part of the LMC-SMC group, and that helps to explain the observed overdensity of globular clusters and satellite dwarfs in the direction of the LMC \citep{Yoon02,Sales17,Erkal20}. Understanding the origin of the LMC-SMC group in the vicinity of the MW has been an active area of research \citep{Cautun19,Evans20,Vasiliev23}.

In this Letter, we report the discovery of a system of interacting dwarf galaxies within a group of at least five confirmed members. The most massive galaxy in this group has a stellar mass of $M_{*}$ = 2.7$\times$10$^{8}$ \msun, which is half of the SMC stellar mass. This Letter is organized as follows. In Section 2, we introduce the system and provide the details of data analysis. In Section 3, we discuss our findings, and we summarize them in Section 4.

\section{Interacting dwarfs in a group environment}

\subsection{Identification}
Our primary focus is to conduct an in-depth analysis of the merging systems of dwarf galaxies in various environments. To achieve this, we have conducted a comprehensive search for such objects by visually inspecting color images of dwarf galaxies in the local volume (z $<$ 0.02) provided by the SDSS and Legacy survey \citep{Aihara11,Dey19}. A comprehensive list of such objects is published in a catalog of merging dwarf galaxies \citep{Paudel18}. In this work, we present a particular case of dwarf galaxy interaction, identifying their unique environment. In particular, we find that dwarf galaxies interact inside a group of dwarf galaxies.

\begin{figure}
\includegraphics[width=8.5cm]{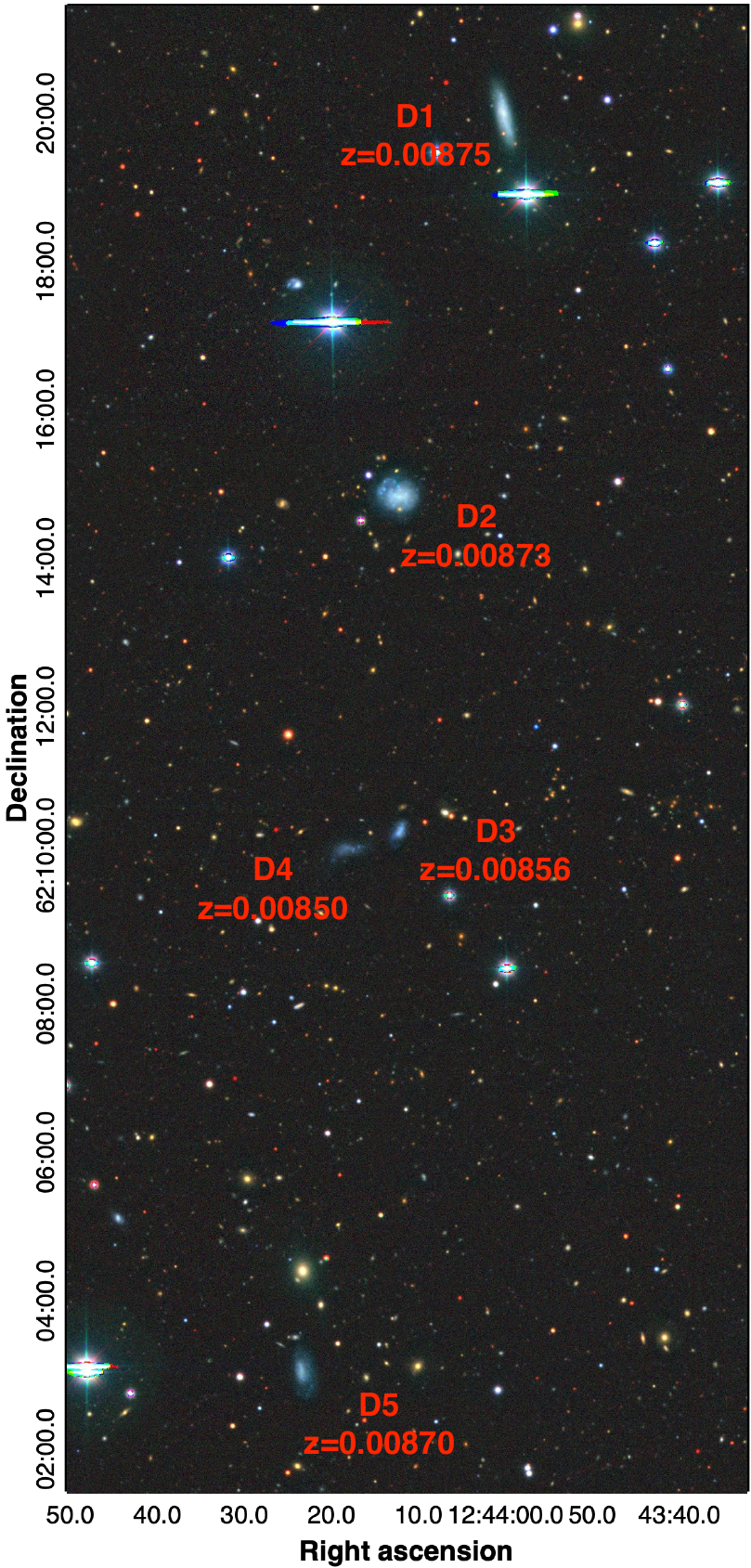}
\caption{The $g-r-z$ combined color image cut-out of the system obtained from the Legacy survey viewer page. The five member dwarf galaxies are designated, and we also provide their redshift information. All five members are arranged in almost a vertical straight line spanning from north to south. The most massive galaxy is a face-on spiral galaxy D2, which has a $B$-band absolute luminosity of $-$16.71 mag.  }
\label{main}
\end{figure}

\begin{table*}
\begin{center}
\caption{Physical Properties of the Member Galaxies}
\begin{tabular}{lcccccccccccccc}
\hline
& R.A. & Decl. &$M_{B}$ &    $v$ &  $R$ &  $g$ & $r$ & $z$ & $g-r$ & $W1$  &   $W2$ &     $M_{*}$ &  Z & $M_{HI}$\\ 
   & hh:mm:ss & dd:mm:ss & (mag) & \kms & kpc & (mag)& (mag) & (mag) & (mag) & (mag) & (mag) & log(\msun) & dex &  log(\msun)  \\
\hline
D1  & 12:43:59 & +62:20:00 & $-$16.42 &    2618  & 49 & 15.97 & 15.45 & 15.25 & 0.52 & 14.76 & 14.58 &  8.29 & 8.36 & 8.57  \\
D2  & 12:44:12 & +62:14:51 & $-$16.71 &    2624  & 00 & 15.71 & 15.28 & 15.18 & 0.43 & 14.56 & 14.47 &  8.44 & 8.33 & 9.02  \\
D3  & 12:44:12 & +62:10:19 & $-$14.74 &    2567  & 43 & 17.74 & 17.50 & 17.50 & 0.24 & 16.29 & 16.07 &  7.65 & 8.23 & 8.62$^{\dagger}$  \\
D4  & 12:44:20 & +62:09:58 & $-$13.98 &    2550  & 44 & 18.49 & 18.22 & 18.30 & 0.27 & 17.31 & 16.99 &  7.17 & 8.00 &   \\
D5  & 12:44:23 & +62:03:06 & $-$15.26 &    2610  & 99 & 17.16 & 16.75 & 16.78 & 0.41 & 16.12 & 16.10 &  7.87 & 8.03 &  $<$\,6.98$^{\ddagger}$ \\
\hline
\end{tabular}
\end{center}
\tablecomments{R is a sky-projected distance of the member galaxies from the most massive member, D2.  A typical error on the magnitudes is 0.01 mag in all-optical band photometry. The stellar masses are derived from infrared band $W1$ and $W1$ magnitudes from the WISE all-sky survey. Z represents the emission line metallicity, 12+log(O/H), which we have derived from the DESI EDR spectra of the central region of galaxies. The total neutral hydrogen masses are derived from the cataloged values of the 21-cm emission line observation of the FASHI survey. $^{\dagger}$Since the FASHI survey uses single-dish observations with a beam size of 2\rlap{.}$\arcmin$9, the interacting pair D3 and D4  are unresolved, and the listed value of HI mass is combined. $^{\ddagger}$D5 is not detected in the FASHI survey; we provide the upper limit of detection based on the survey depth; see the text for details.}
\label{aphot}
\end{table*}

We have identified a unique group of five star-forming dwarf galaxies located in a near isolated environment at a distance of $D$ = 36 Mpc\footnote{Based on Hubble flow with a mean redshift of the group z= 0.0086}; We find no massive galaxies ($M_{*}$ $>$ 10$^{10}$ \msun) within 700 kpc sky projected radius and line-of-sight radial velocity range of $\pm$700 \kms{} as we have searched in Nasa Extragalactic Database\footnote{https://ned.ipac.caltech.edu/}. These values typically correspond to twice the projected viral radius and velocity dispersion of a group with a central galaxy of stellar mass $\sim$10$^{10}$ \msun, respectively \citep{Makarov11}.

In Figure \ref{main}, we show the color image cut-out of the system, which we have obtained from the Legacy survey viewer\footnote{https://www.legacysurvey.org/viewer}. We see that member dwarf galaxies are arranged in almost a vertical straight line spanning from north to south, where the most massive galaxy is a face-on spiral, D2, which has a $B$-band absolute luminosity of $-$16.71 mag (similar to that of the SMC, $M_{B}$ = $-$16.79 mag). As the color image reveals, all member galaxies are blue and star-forming. The spatial span of the group (largest sky-projected separation between member galaxies) is 154  kpc, and the velocity dispersion is 33 \kms.  A Principal component analysis (PCA) was employed to fit a planar structure, yielding a semi-major axis of 57 kpc and a semi-minor axis of 4 kpc, which results in an axis ratio of 0.06.

In the vicinity of the most massive galaxy of the group, D2, we identify a pair of dwarf galaxies (D3 and D4) located at a sky-projection distance of 10.3 kpc southward with a relative line-of-sight velocity of 30 \kms. A careful visual inspection of the optical images\footnote{To gain signal in the low-surface brightness region, tidal tails, we coadded the Legacy g-r-z band images.} of the pair reveals their disturbed morphology, showing a sign of tidal interaction where the tidal tails are elongated in the opposite direction of the interaction. The relative line-of-sight velocity between the pair is 17 \kms.  The $B$-band absolute luminosities of D3 and D4 are $-$14.74 and $-$13.98 mag, respectively, and their stellar mass ratio is 3.

\subsection{Data Analysis}
This work benefits from the substantial multi-wavelength data available in public archives, which allowed us to perform the required detailed measurements of the physical properties of the group and its member galaxies. Particularly, Dark Energy Spectroscopic Instrument (DESI\footnote{https://data.desi.lbl.gov/}) early data release \citep[EDR][]{DESI24} spectroscopic data enabled us to confirm its group member with radial velocity measurement. We used the Legacy survey imaging data to calculate optical $g$, $r$, and $z$-band magnitudes and  Strasbourg Astronomical Data Center (CDS\footnote{https://cds.unistra.fr}) catalog services to fetch other physical properties of the member galaxies.

We downloaded $g$, $r$, and $z$-band images from the Legacy survey database, and the observation was made through the  Beijing-Arizona Sky Survey \citep[BASS,][]{Zou17}. We performed aperture photometry of member galaxy using the similar method we had implemented in our main catalog paper \citep{Paudel18}. Finally, the $B$-band magnitudes were derived from the $g$ and $r$ magnitudes using the filter conversion equation, $B=g+0.227(g-r) -0.337$\footnote{https://www.sdss3.org/dr8/algorithms/sdssUBVRITransform.php}.  The $B$-band magnitude of member galaxies ranges from $-$16.71 to $-$13.98 mag, the former being similar to the SMC and the latter being similar to the Fornax dwarf spheroidal galaxy. As expected for star-forming dwarf galaxies, the member dwarf galaxies have $g-r$ indices ranging from 0.2 to 0.5 mag.

\begin{figure}
\includegraphics[width=8.5cm]{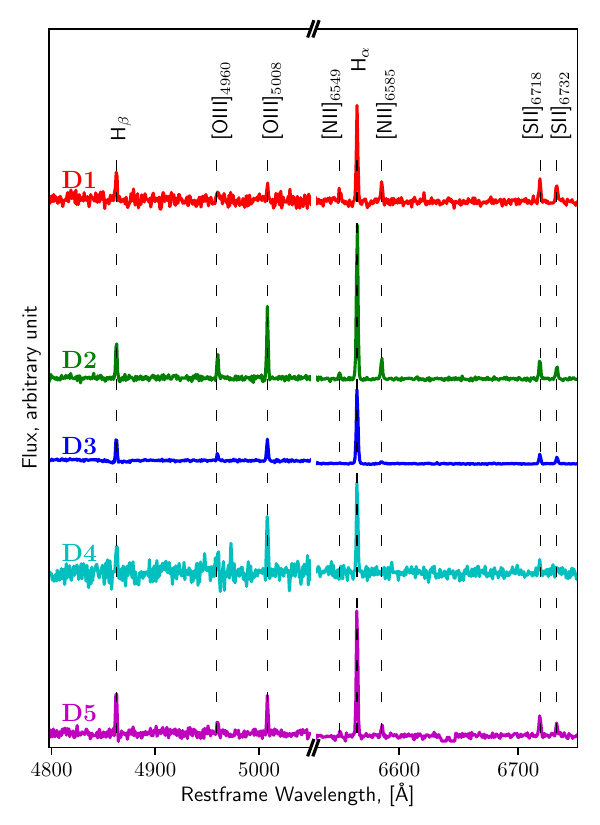}
\caption{The DESI EDR optical spectra of the central region of member galaxies. The observed spectra are shifted to the rest frame wavelength and smoothed with a three-pixel Gaussian kernel. The major emission lines are marked with vertical dash lines.}
\label{dspec}
\end{figure}

We obtained infrared band $W1$ and $W2$ magnitudes from the Wide-field Infrared Survey Explorer (WISE; \citealt{Wright10})  all-sky survey catalog \citep{Cutri14} and derived the stellar mass using an empirical formula from \cite{Eskew12}. The brightest central galaxy has a stellar mass of 2.75 $\times$ 10$^{8}$ \msun, which is nearly half of the SMC stellar mass \citep{McConnachie12}. The least massive member galaxy, D4, has a stellar mass of 1.47 $\times$ 10$^{7}$ \msun, similar to that of Fornax dwarf spheroidal, and the median of stellar masses of the five member galaxies is 7.87 $\times$ 10$^{7}$ \msun.

Using the Five-hundred-meter Aperture Spherical Radio Telescope (FAST\footnote{https://fast.bao.ac.cn}), the FAST All Sky HI survey (FASHI; \citealt{Zhang24}) has covered the sky around our group. In our galaxy group region, three sources are detected in 21-cm emission line observation with a radial velocity similar to the member galaxies: D1, D2, and D3. For the D5, there is no 21-cm radio emission detection in FASHI, and for the D3 and D4, the detection is combined, likely due to the separation between them being smaller than FASHI 2\rlap{.}$\arcmin$9 beam size. We derive total neutral hydrogen (HI) mass from 21-cm emission fluxes using a formula: \\
$M_{HI} (M_{\sun})= 2.36 \times 10^{5} D^{2} \int F dV$, where $\int F dV$ is integrated flux in Jy-\kms,  and $D$ is distance to galaxy in Mpc. For the undetected galaxy, D5, we estimated the limit of detection using the formula from \cite{Gavazzi08}:  \\
$M_{HI,lim}(M_{\sun}) = SNR \times rms \times W_{50} \times D^{2} \times 2.36\times10^{5}$, 
with the minimum SNR of 1 and $W_{50}$ = 50 \kms of our dwarf galaxies and the median noise $rms$ of the survey as $rms$ = 1.5 mJy, and the distance $D$ = 36 Mpc.

Fortunately, the sky region is also covered by DESI EDR. We identify all member galaxies targeted by the DESI spectroscopic observation, including multiple observations for D1, D2, and D5. Indeed, this helps us to confirm the membership of these galaxies with radial velocity measurement. We find that the overall range of radial velocity of the member galaxies ranges from 2550 to 2624 \kms\, with a standard deviation $\sigma_{v}$ = 33 \kms. The optical spectra of these galaxies strongly show H$_{\alpha}$ emission, as shown in Figure \ref{dspec}.  We estimated the oxygen abundance, 12+log(O/H), using a combination of line ratios H$_{\alpha}$/[NII] and [OIII]/ H$_{\beta}$ \citep{Marino13}. We list the photometric and physical parameters of member galaxies in Table \ref{aphot}.

\begin{figure}
\includegraphics[width=8.5cm]{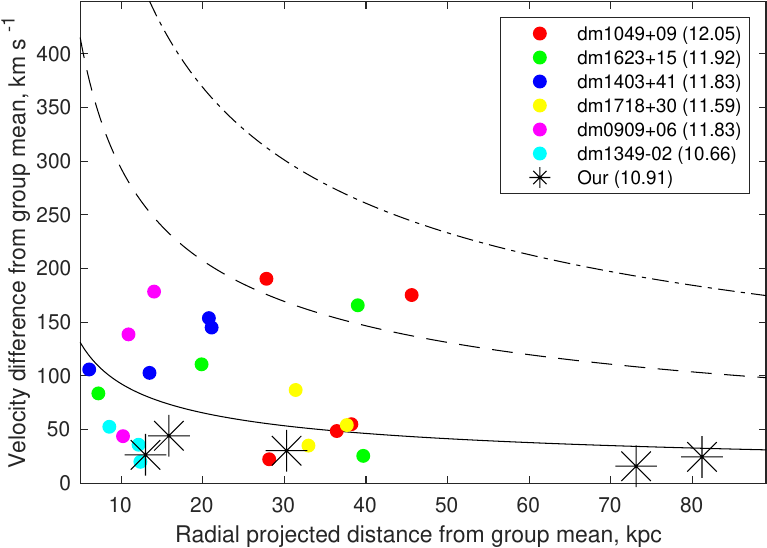}
\caption{The position of the member galaxies (black asterisk) in a phase-space diagram.  For comparison, we use the S17 groups, where different colors represent different groups and their respective group dynamical mass (Log(M$_{dyn}$/\msun)) are listed in the legend. The solid, dashed, and dash dotted line represents the escape velocity curve for a halo mass of 10$^{10}$, 10$^{11}$ and 10$^{11.5}$\msun, respectively. 
}
\label{phase}
\end{figure}

The total dynamical mass of the system is calculated as the total projected mass of the group  \citep{Heisler85} using an equation:\\
$M =\frac{32}{\pi}\frac{1}{G(N-3/2)} \sum\limits_{i}^{N} R_{p,i} \Delta v_{i}^{2}$,\\
where v$_{i}$ and R$_{p,i}$ are, respectively, radial velocity and projected distance of the $i$th galaxy relative to the system's center. $N$ is the total number of the members of galaxies. Using this equation, we obtained the total dynamical mass of the system $M_{dyn}$ = 6.02$\times$10$^{10}$ M$_{\sun}$ whereas the total baryonic mass (stellar + HI) of the group is 2.6$\times$10$^{9}$ \msun, which gives us the dynamical to baryonic mass ratio of 23.

The 3D velocity dispersion ($\sigma_{3D}$) is calculated using line-of-sight velocities of all the group members and corrected for potential transverse motion using an equation: $\sigma_{3D}$ = $\sqrt{3}$ $\times$ $\sqrt{\langle v^{2} \rangle - \langle v \rangle^{2}}$, where $v$ is line-of-sight velocities. We obtained the value of $\sigma_{3D}$  = 51 \kms. Finally, we calculated the minimum mass ratio required for the group to be a gravitationally bound structure. For this, we used the escape velocity of the group equal to $\sigma_{3D}$ and found that the mass required for the bound group is 2.99$\times$10$^{10}$ \msun, which is half of the total dynamical mass and approximately ten times of total Baryonic mass. 


\section{Discussion}
In this work, we presented a unique group of five dwarf galaxies in a relatively isolated environment. All member galaxies of the group are star-forming and blue. Despite searching the SDSS and DESI EDR spectroscopic databases within a 500 kpc radius and a relative line-of-sight velocity range of $\pm$500 \kms, no additional members were identified. The SDSS has targeted D1, D2, and D5, while the DESI has targeted all five members. Indeed, this may give a biased view, potentially overlooking non-star-forming dwarf galaxies. Notably, measuring the redshift of star-forming dwarf galaxies is considerably easier and more effective due to the presence of emission lines compared to non-star-forming dwarf galaxies characterized by absorption-dominated spectra. It is important to note that this analysis is based solely on the five confirmed star-forming dwarf galaxy members. There is a possibility that additional member dwarfs could be discovered through a dedicated, in-depth spectroscopic survey of the area.

\begin{figure}
\includegraphics[width=8.5cm]{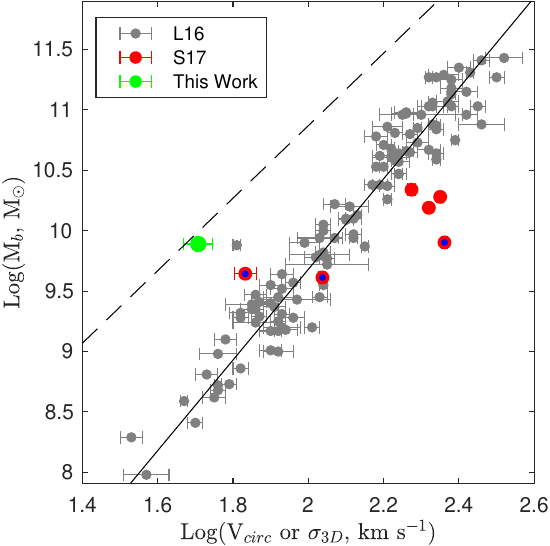}
\caption{The baryonic Tully-Fisher relation plotted as total baryonic mass against $\sigma_{3D}$ of dwarf galaxy groups. For comparison, we plot the total baryonic mass against the rotation velocity of galaxies from L16. The S17 sample is shown by red circles, where the blue dots represent the triplet group. The best-fit value of L16 data shown by a solid line and the dashed line is derived from cosmic baryon fraction using an equation $M_{200}f_{b} \propto M_{200}^{3}f_{V}^{3}$, with a value of $f_{b}$ = 0.17 and $f_{V}$ = 1.
Our group (green circles) stands out, being a significant outlier on the left side of the BTFR, and nearly falls along the line of the cosmic average baryon fraction.
} 
\label{tflr}
\end{figure}

\subsection{Comparison to Previous Work}
 
\citealt{Stierwalt17} (S17) studied seven isolated compact groups of galaxies of redshift range from 0.02 to 0.05 ($D$ $\simeq$ 80 to 200 Mpc). Compared to the S17 groups, our group represents a much closer system at $D$ = 36 Mpc. In the S17 sample, only one group has five members, which is also the most massive group with a total dynamical mass of 1.12 $\times$10$^{12}$ \msun, and there are two quads and three triplets. Among the three triplets, the group 1440+14 is located near a relatively bright ($M_{B}$ $<$ $-$19 mag) galaxy (at a distance of 230 kpc with a radial velocity separation of 90 \kms). We, therefore, drop this group from our comparison sample. In Figure \ref{phase}, we show a phase-space diagram and compare our group to the S17 groups, where the solid and dash lines represent the escape velocity curve for a halo mass of 10$^{10}$  and 10$^{11}$\msun, respectively. It is interesting to note that the member galaxies of our group fall well below the escape velocity curve for a halo mass of 10$^{10}$ \msun, confirming the bound structure.  In fact, a few members of the most massive group of S17 are outside of the escape velocity curve of the halo mass of 10$^{11}$ \msun  (red and green dots), but reported dynamical masses of these groups are on the order of 10$^{12}$ \msun. One main difference between the S17 groups and our group is that the S17 groups are compact, with the largest group having a size of 80.5 kpc, whereas our group has 154 kpc. This is probably due to the selection criteria of S17: they deliberately selected a sample of compact groups. On the other hand, S17 groups have a significantly larger velocity of dispersion compared to our group. The largest S17 group (which also has a five-member) has a velocity of dispersion of $>$130 \kms{} whereas our group has a velocity dispersion of 33 \kms.

In Figure \ref{tflr}, we show the baryonic Tully-Fisher Relation (BTFR) plotted as total baryonic mass against $\sigma_{3D}$ of dwarf galaxy groups. For comparison, we plotted a well-studied sample of galaxies of  \citealt{Lelli16} (L16). It is evident that the majority of S17 groups fall on the right side of the BTFR defined by L16 galaxies, indicating an excess of baryon mass in these systems. We also highlighted the triplet groups of S17 (with blue dots), which have nearly similar total baryon mass. On the other hand, our group stands out, being a significant outlier on the left side of the BTFR; in fact, it nearly falls along the line of the cosmic average baryon fraction, indicating that there are no missing baryons in this group, and all baryons from the Big Bang nucleosynthesis are locked inside its member dwarf galaxies.

\subsection{Uniqueness, Chance or Real?}
The ongoing interaction between the LMC and SMC is indeed happening in a group environment in the vicinity of the MW \citep{Putman98}. However, satellite pair interaction in a group environment is rare \citep{Robotham12,Paudel17,Paudel20}. In our merging dwarf galaxy catalog, only five percent of dwarf-dwarf galaxy interactions happen in group environments. In our group of this study, we also find that the dwarf galaxies, D3 and D4, of stellar masses similar to Fornax dwarf spheroidal, are interacting in the vicinity of the group's most massive host. The LMC-SMC pair has been studied in great detail and is regarded as an essential laboratory for studying the evolution of dwarf galaxies and the formation of anisotropic distribution of satellite galaxies around their host. Recent analyses have shown that the LMC-SMC pair is not only interacting but also constitutes a group of dwarf galaxies, which include at least five other MW satellites \citep{Zhang19,Patel20}.

\begin{figure}
\includegraphics[width=9cm]{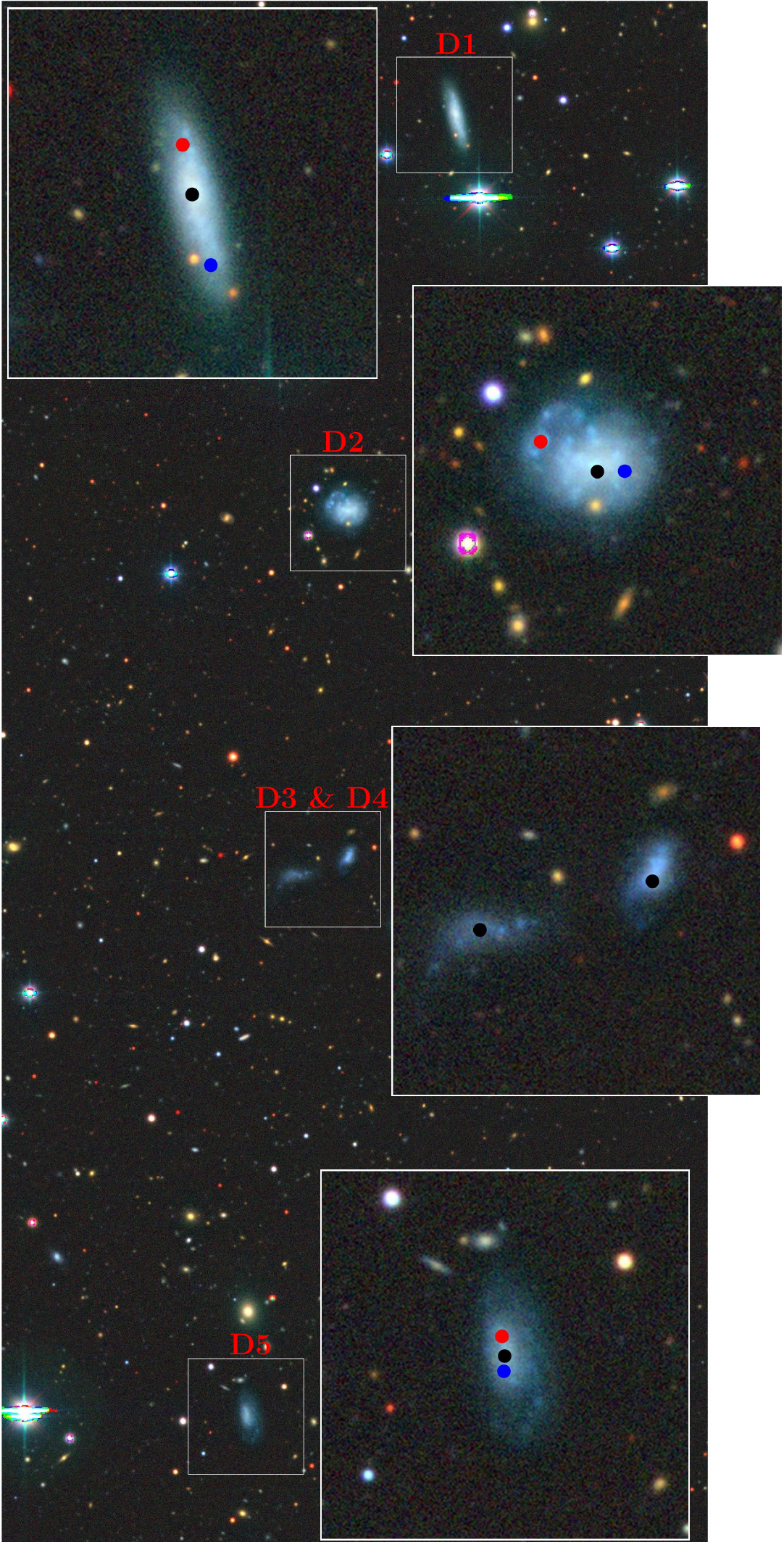}
\caption{A zoomed-in view of the member galaxies. The 1$\arcmin$$\times$1$\arcmin$ square box is zoomed to a better view, where we marked the central positions with the black dots. The red and blue dots represent the red-shifted and blue-shifted parts of the galaxy with respect to their center, respectively.  }
\label{zmfig}
\end{figure}

One of the unique features of our group is the near straight-line distribution of the member galaxies in sky projection. To confirm whether this is a chance projection or real is beyond the observational limit of this work.  In particular, we do not know the distances of individual galaxies from which we could make a three-dimensional picture of the group. In fact, the satellites around our Milky Way galaxy are distributed in an extremely thin plane with an axis ratio of 0.29 \citep{Pawlowski12,Pawlowski14}, and there are a dozen similar structures have been identified in various galaxy groups \citep{Libeskind19}. Extending this phenomenon, we may have observed similar structures on a much smaller scale in contrast to previously discovered planar satellite systems around the massive host with a stellar mass of $\sim$10$^{11}$ \msun.

In addition to their planar distribution, we also find that the member galaxies in this group have similar rotational directions.  In Figure \ref{zmfig}, we show a zoom-in view of the member galaxies, where the relative line-of-sight velocity directions with respect to the galaxy centers are marked with blue and red symbols for redshifted and blueshifted velocities, respectively. Thanks to the DESI, the EDR provides multiple spectroscopic observations for D1, D2, and D5 from different parts of galaxies, which allow us to tentatively estimate their direction of rotation. D1 and D5 are fairly edge-on galaxies, and with the radial velocity measurement from different ends of galaxies, we can clearly determine their rotational direction.
D2, on the other hand, has a nearly face-on orientation. While we can still estimate its rotational direction using spatially resolved spectroscopy, the data statistics are, unfortunately, quite limited. We noticed that  D1, D2, and D5 have a common rotational direction in that the north parts of these galaxies are red-shifted, and the lower south parts are blue-shifted.  

\section{Summary}
In this work, we have studied a rare group of dwarf galaxies consisting of at least five confirmed members, two of which are interacting. The member galaxies are star-forming and blue. The brightest member galaxy has a stellar mass of 2.75 $\times$ 10$^{8}$ \msun, which is nearly half of SMC's stellar mass. The median stellar mass of group members is 7.87 $\times$ 10$^{7}$ \msun.  We find that all galaxies in the group are arranged nearly in a straight line, forming an exceptionally thin planar structure. The observed spatial span of the member galaxy (distance from one end to another) is 154 kpc, and the maximum difference in line-fo-sight velocity between them is 75 \kms. 
The derived total dynamical of the system is $M_{\rm dyn}$ = 6.02$\times$10$^{10}$ \msun, whereas the total baryonic mass (stellar + HI) of the group is 2.6$\times$10$^{9}$ M$_{\sun}$, which gives us the dynamical to baryonic mass ratio of 23. Using $\sigma_{3D}$, we derived the minimum mass ratio required for the group to be a gravitationally bound structure, which comes out to be 2.99$\times$10$^{10}$ \msun.  We also found that our group is a significant outlier from the standard BTFR defined by galaxies, indicating that all baryons of the group are locked inside the member galaxies.

We have identified a distinct group of dwarf galaxies, with all five members aligned along a straight line in the celestial plane and three sharing a common rotational direction. The significance of the observed features and whether they are coincidental or genuine remains uncertain.
It has been argued that observations of coherent structures like a plane of satellites or rotating satellite systems are against our well-known $\Lambda$CDM cosmology, which assumes the hierarchical build-up of large-scale structures \citep{Libeskind15,Pawlowski17,Pawlowski21}. Further investigation of this system, both observationally and theoretically, could shed light on the prevalence of such configurations in the $\Lambda$CDM Universe and potentially reveal the accuracy of our current understanding of large-scale structure formation.
Moreover, subsequent high-resolution 21-cm radio observations may provide detailed insights into the internal kinematics of each dwarf galaxy, potentially confirming their individual rotational characteristics.

\newpage
\acknowledgments
SP and SJY acknowledge support from the Mid-career Researcher Program (RS-2023-00208957 and RS-2024-00344283, respectively) through Korea's National Research Foundation (NRF). 
SJY and CGS acknowledge support from the Basic Science Research Program (2022R1A6A1A03053472 and 2018R1A6A1A06024977, respectively) through Korea's NRF funded by the Ministry of Education. J.Y. was supported by a KIAS Individual Grant (QP089902) via the Quantum Universe Center at Korea Institute for Advanced Study

The DESI Legacy Imaging Surveys consist of three individual and complementary projects: the Dark Energy Camera Legacy Survey (DECaLS), the Beijing-Arizona Sky Survey (BASS), and the Mayall z-band Legacy Survey (MzLS). DECaLS, BASS and MzLS together include data obtained, respectively, at the Blanco telescope, Cerro Tololo Inter-American Observatory, NSF’s NOIRLab; the Bok telescope, Steward Observatory, University of Arizona; and the Mayall telescope, Kitt Peak National Observatory, NOIRLab. NOIRLab is operated by the Association of Universities for Research in Astronomy (AURA) under a cooperative agreement with the National Science Foundation. Pipeline processing and analyses of the data were supported by NOIRLab and the Lawrence Berkeley National Laboratory (LBNL). Legacy Surveys also uses data products from the Near-Earth Object Wide-field Infrared Survey Explorer (NEOWISE), a project of the Jet Propulsion Laboratory/California Institute of Technology, funded by the National Aeronautics and Space Administration. Legacy Surveys was supported by: the Director, Office of Science, Office of High Energy Physics of the U.S. Department of Energy; the National Energy Research Scientific Computing Center, a DOE Office of Science User Facility; the U.S. National Science Foundation, Division of Astronomical Sciences; the National Astronomical Observatories of China, the Chinese Academy of Sciences and the Chinese National Natural Science Foundation. LBNL is managed by the Regents of the University of California under contract to the U.S. Department of Energy. The complete acknowledgments can be found at https://www.legacysurvey.org/acknowledgment/.

This research used data obtained with the Dark Energy Spectroscopic Instrument (DESI). DESI construction and operations is managed by the Lawrence Berkeley National Laboratory. This material is based upon work supported by the U.S. Department of Energy, Office of Science, Office of High-Energy Physics, under Contract No. DE–AC02–05CH11231, and by the National Energy Research Scientific Computing Center, a DOE Office of Science User Facility under the same contract. Additional support for DESI was provided by the U.S. National Science Foundation (NSF), Division of Astronomical Sciences under Contract No. AST-0950945 to the NSF’s National Optical-Infrared Astronomy Research Laboratory; the Science and Technology Facilities Council of the United Kingdom; the Gordon and Betty Moore Foundation; the Heising-Simons Foundation; the French Alternative Energies and Atomic Energy Commission (CEA); the National Council of Science and Technology of Mexico (CONACYT); the Ministry of Science and Innovation of Spain (MICINN), and by the DESI Member Institutions: www.desi.lbl.gov/collaborating-institutions. The DESI collaboration is honored to be permitted to conduct scientific research on Iolkam Du’ag (Kitt Peak), a mountain with particular significance to the Tohono O’odham Nation. Any opinions, findings, and conclusions or recommendations expressed in this material are those of the author(s) and do not necessarily reflect the views of the U.S. National Science Foundation, the U.S. Department of Energy, or any of the listed funding agencies.

\vspace{5mm}


\end{document}